\documentclass[10pt,letterpaper,twocolumn,showpacs,pra,aps]{revtex4-1}

\usepackage{amsmath}  
\usepackage{amsfonts} 
\usepackage{graphicx} 
\usepackage{amssymb}

\begin{document}

\title{Sarma-Bogomol'nyi equations in superconductivity}

\author{Mia Kyler and Eugene B. Kolomeisky}

\affiliation
{Department of Physics, University of Virginia, P. O. Box 400714, Charlottesville, Virginia 22904-4714, USA}

\date{\today}

\begin{abstract}
Topological defects occurring in nonlinear classical field theories are described by a system of second-order differential equations.  A breakthrough was made in 1976 by E. B. Bogomoln'yi who demonstrated that in several field theories these equations can be reduced to first-order provided the coupling constants take on particular values.  One of the examples involved a string in the Abelian Higgs model which is equivalent to the Abrikosov flux line of the Ginzburg-Landau theory of superconductivity.  In a similar vein, in the 1966 textbook \textit{Superconductivity of Metals and Alloys} P. G. de Gennes explained how to reduce the second-order Ginzburg-Landau equations to first-order at a particular value of the Ginzburg-Landau parameter by a method due to G. Sarma.  We analyze the two ways of arriving at the first-order Sarma-Bogomol'nyi equations and conclude that while they both rely on the same operator identity, Sarma's method is free of the assumption that there is a topological defect.  The implication is that Bogomol'nyi equations found in other field theories may be a source of a wider range of solutions beyond topological defects.

\end{abstract}


\maketitle

\section{Introduction}

Second-order differential equations are ubiquitous in physics, but occasionally one encounters special cases where the order of these equations may be reduced.  One example of such a reduction is the quantum harmonic oscillator as solved by Dirac.  Dirac introduced the annihilation and creation operators, which permit solving of the second-order problem posed by the linear Schr\"odinger equation by reduction to a first-order linear equation \cite{Dirac}.   

Topological defects - domain walls, vortices, strings and monopoles - arising in nonlinear classical field theories are described by second-order nonlinear equations.  A promising direction in the study of topological defects was found by E. B. Bogomol'nyi \cite{Bogo}, who noticed that in several field theories these equations can be reduced to first-order nonlinear equations with appropriate choice of the coupling constants.  In these theories defect energies are bounded from below by a quantity linear in the magnitude of a topological quantum number \cite{BPST};  the equality is realized when the field configuration satisfies the first-order Bogomol'nyi equations.  Since then many more examples of such reductions and behavior were found \cite{Manton,Weinberg}.

One of Bogomol'nyi's examples was a string in the Abelian Higgs model \cite{string,Vega} known to be equivalent to the Abrikosov flux (vortex) line \cite{Abrikosov} of the Ginzburg-Landau (GL) theory of superconductivity \cite{GL,LL9}.  The underlying theory in both cases is that of the complex scalar field coupled to a gauge field.  Curiously, in 1966 textbook \cite{de Gennes} P. G. de Gennes explained that at the value of the GL parameter $\kappa=1/\sqrt{2}$, which separates type-I ($\kappa<1/\sqrt{2}$) from type-II ($\kappa>1/\sqrt{2}$) superconductors \cite{Abrikosov,GL}, the second-order nonlinear GL equations can be reduced to first-order nonlinear equations.  The latter were then employed to show that the surface energy between the normal and superconductive phases vanishes at $\kappa=1/\sqrt{2}$.  De Gennes credited G. Sarma with the method used to arrive at this conclusion.  The method consists in the introduction of operators analogous to Dirac's creation and annihilation operators \cite{Dirac} and then employing an operator identity to reduce the order of the GL equations.  Additionally, de Gennes claimed that ``a similar calculation in terms of Sarma solutions can be done for the energy of an isolated vortex line, when $\kappa=1/\sqrt{2}$" \cite{de Gennes}.

While acknowledging \cite{Bogo} that the same equations were ``also obtained by means of the substitution used in the theory of superconductivity for solving some problems at $\kappa=1/\sqrt{2}$", no comparative analysis was given, and reference to the textbook \cite{de Gennes} was generic. 

To the best of our knowledge Sarma's contribution is virtually unknown.  The exception is a recent work on superconductors at $\kappa=1/\sqrt{2}$ \cite{recent1,recent2,recent3} that does mention ``Sarma solution" \cite{recent1}.  These authors exploit degeneracy of the ground state at $\kappa=1/\sqrt{2}$ \cite{Manton,Weinberg} to show the existence of a variety of exotic solutions to the GL equations.  While the reference to the textbook \cite{de Gennes} remains generic, an effort was made to describe Sarma's method \cite{recent1,recent3}.  However it is not quite the same as the original \cite{de Gennes}.

The goal of this paper is to provide detailed analysis of both the Sarma and Bogomol'nyi methods to reduce second-order nonlinear GL equations to first-order nonlinear Sarma-Bogomol'nyi (SB) equations.  In a nutshell, the difference between the two methods can be traced to whether one is working with the GL field equations (Sarma) or the GL free energy functional (Bogomol'nyi).  Our conclusion is that Sarma's method is more general as it does not rely on the existence of a topological defect.  The implication is that Bogomol'nyi equations found in other field theories may apply more generally and serve as a source of new solutions beyond topological defects.

The rest of this paper is organized as follows.  We begin with outlining the GL theory (Section II) followed by a description of Abrikosov's solution for the flux line (Section III).  In the spirit of Bogomol'nyi's method, we then show (Section IV) how Abrikosov's equations can be reduced to the first order SB equations at the value of the GL parameter $\kappa=1/\sqrt{2}$. This analysis employs the cylindrical symmetry of the solution combined with the fact that the vortex line carries quantized magnetic flux. Then (Section V), we employ Sarma's operator method to reduce the second order GL equations to the first order SB equations in their most general form and illustrate that in the particular case of cylindrical symmetry, they recover the first order equations of the previous section. Finally (Section VI), we show a procedure that combines the Bogomol'nyi and Sarma methods to derive the general SB equations without the assumption of cylindrical symmetry. We conclude with a summary of our results and possible implications for other field theories.

\section{Statement of the problem}

Our starting point is the GL free energy functional \cite{GL,LL9} below the point of the superconductive transition   
\begin{eqnarray}
\label{GL_functional}
\mathcal{F}&=&\int \Bigg\{ \frac{\hbar^{2}}{2m} \left | \left (\mathbf{\nabla} - \frac{ie}{\hbar c} \mathbf{A}\right) \Psi \right |^{2}
+\frac{ms^{2}}{2n_{0}}(|\Psi|^{2}-n_{0})^{2}\nonumber\\
&+&\frac{\mathbf{H}^{2}}{8\pi} \Bigg\} dV,~~~~~\mathbf{H}=\nabla\times\mathbf{A}
\end{eqnarray}
where $m$ is the effective mass of the Cooper pair of net charge $e$, $\mathbf{A}$ is the vector potential, $\mathbf{H}$ is the magnetic field, and $\Psi$ is the macroscopic wave function.  The second term of the integrand represents the free energy density cost of uniform deviation of the number density of superconductive electrons $|\Psi|^{2}$ from equilibrium value of $n_{0}$, and the form of temperature-dependent coefficient $ms^{2}/2n_{0}$ is chosen for aesthetic reasons;  the quantity $s$ has dimensionality of velocity.  The energy density of the magnetic field is given by the last integrand in Eq.(\ref{GL_functional}).    

Minimizing the free energy functional (\ref{GL_functional}) with respect to $\Psi^{*}$ and $\mathbf{A}$ one arrives at the GL equations \cite{GL,LL9}:
\begin{equation}
\label{GL_equation}
\left[ -\frac{\hbar^2}{2m} \left( \nabla - \frac{ie}{\hbar c}\mathbf{A}\right)^2 + \frac{ms^2}{n_0}(|\Psi|^2-n_0)\right]\Psi = 0,
\end{equation}
\begin{equation}
\label{Ampere's_law}
\nabla \times \mathbf{H} = \frac{4\pi}{c}\mathbf{j},~~\mathbf{j} = \frac{ieh}{2m}(\Psi\nabla\Psi^{*} - \Psi^{*}\nabla\Psi)-\frac{e^2}{mc}|\Psi|^2\mathbf{A}
\end{equation}
where the second equation is Amper\`e's law and $\mathbf{j}$ is the current density.  Below we will be often using the Madelung representation of the wave function \cite{Madelung}
\begin{equation}
\label{Madelung}
\Psi(\mathbf{r})=\sqrt{n(\mathbf{r})}e^{i\theta(\mathbf{r})}
\end{equation}  
where $n=|\Psi|^{2}$ is the number density and $\theta$ is the phase of the wave function.  This transforms the free energy functional (\ref{GL_functional}) into an equivalent form
\begin{equation}
\label{GL_Madelung}
\mathcal{F}= \int \Bigg\{\frac{\hbar^2}{2m}(\nabla\sqrt{n})^{2} + \frac{mn\mathbf{v}^{2}}{2} +\frac{ms^{2}}{2n_{0}}(n-n_{0})^{2} + \frac{\mathbf{H}^2}{8\pi}\Bigg\}dV
\end{equation} 
where
\begin{equation}
\label{flow_velocity}
\mathbf{v} = \frac{\hbar}{m} \left(\mathbf{\nabla}\theta-\frac{e}{\hbar c}\mathbf{A}\right)
\end{equation}
is the superconductive flow velocity.  Then the second integrand in Eq.(\ref{GL_Madelung}) is the kinetic energy density of the flow while the first is the energy density cost of having inhomogeneous $n$.  Minimizing the free energy functional (\ref{GL_Madelung}) with respect to the density $n$, phase $\theta$, and vector potential $\mathbf{A}$ or, equivalently, substituting the wave function (\ref{Madelung}) in Eqs.(\ref{GL_equation}) and (\ref{Ampere's_law}) gives the GL equations in the Madelung representation
\begin{equation}
\label{GL_equation_Madelung}
\frac{m\mathbf{v}^{2}}{2}+\frac{ms^{2}}{n_{0}}(n-n_{0})-\frac{\hbar^{2}}{2m}\frac{\nabla^{2}\sqrt{n}}{\sqrt{n}}=0,
\end{equation}
\begin{equation}
\label{stationary_continuity}
\nabla \cdot (n\mathbf{v}) = 0,
\end{equation}
\begin{equation}
\label{Ampere's_law_Madelung}
\nabla \times \mathbf{H} = \frac{4\pi}{c}en\mathbf{v}
\end{equation}
where Eq.(\ref{stationary_continuity}) is the stationary version of the continuity equation.  

The two relevant length scales of the problem, the coherence length, $\xi$ and the London penetration depth $\lambda$, as well as their ratio $\kappa$, the GL parameter, are defined as
\begin{equation}
\label{parameters}
\xi^2 = \frac{\hbar^2}{2m^2s^2},~~\lambda^2 = \frac{mc^{2}}{4\pi n_{0}e^{2}},~~\kappa = \frac{\lambda}{\xi}.
\end{equation}   

\section{Abrikosov equations}

One of the hallmarks of the superconductive state is the existence of topological defects, the Abrikosov flux lines \cite{Abrikosov}, that play a central role in the destruction of superconductivity by a magnetic field in type-II superconductors.  An individual flux line possesses the following properties \cite{Abrikosov,LL9}:  

The superconductive density $n(\rho)$ (where $\rho$ is the distance from the vortex axis coinciding with the $z$-axis of cylindrical system of coordinates ($\rho$, $\varphi$, $z$)) is suppressed within the vortex core, whose size is set by the coherence length $\xi$ (\ref{parameters}).  The $n(\rho)$ dependence is an increasing function of $\rho$ that approaches its bulk value $n(\rho\rightarrow\infty)=n_{0}$ from below.  There is an accompanied profile of the magnetic field pointing in the $z$ direction;  the field is a decreasing function of $\rho$ falling off over a length scale set by the London penetration depth $\lambda$ (\ref{parameters}).  The magnetic flux associated with the flux line is quantized in units of the flux quantum $2\pi\hbar c/|e|$.

The flux line is a $z$-independent solution to the GL equations that minimizes the free energy (\ref{GL_Madelung}) per unit length.  Since the integrand is a sum of positively-defined contributions, the necessary condition for the minimum is that every one of its integrands vanishes as $\rho\rightarrow\infty$.  

When applied to the first two terms involving $n(\rho,\varphi)$, this criterion implies that $n(\rho,\varphi)\rightarrow n_{0}$ as $\rho\rightarrow\infty$, and the wave function (\ref{Madelung}) has the limiting behavior of the form $\Psi(\rho\rightarrow\infty,\varphi)=\sqrt{n_{0}}\exp[ i\theta(\varphi)]$.  The constraint that it must be single-valued then dictates that      
\begin{equation}
\label{single-valuedness}
\theta(\varphi+2\pi)-\theta(\varphi)=2\pi l,~~~l=0,\pm 1,\pm 2,...
\end{equation}
The requirement that $mn\mathbf{v}^{2}/2$, the kinetic energy density term in (\ref{GL_Madelung}), vanishes as $\rho\rightarrow\infty$ combined with the definition of the flow velocity (\ref{flow_velocity}) implies that
\begin{equation}
\label{A_large_rho}
\mathbf{A}=\frac{\hbar c}{e}\nabla \theta=\frac{\hbar c}{e}\frac{1}{\rho}\frac{d\theta}{d \varphi}\mathbf{e}_{\varphi}, ~\text{as}~\rho\rightarrow\infty.
\end{equation}
(hereafter $\mathbf{e}_{\rho,\varphi,z}$ are unit vectors of the cylindrical system of coordinates).  As a result the magnetic field $\mathbf{H}=\nabla\times\mathbf{A}$ vanishes as $\rho\rightarrow\infty$ as well.

Eqs.(\ref{single-valuedness}) and (\ref{A_large_rho}) also imply that the magnetic flux over the entire $xy$ plane is quantized in units of $2\pi\hbar c/|e|$, the flux quantum \cite{LL9}:   
\begin{equation}
\label{flux_quantization}
\int \mathbf{H}\cdot d\mathbf{s}=\oint \mathbf{A}\cdot d\mathbf{l}
=\frac{\hbar c}{e}\oint \nabla\theta\cdot d\mathbf{l}=\frac{2\pi\hbar c}{e}l
\end{equation}
where $d\mathbf{s}=\mathbf{e}_{z}dxdy$, $d\mathbf{l}$ is an infinitesimal element of a contour enclosing the $xy$ plane and Stokes's theorem was employed in the second step.  

In the same $\rho\rightarrow\infty$ limit the continuity equation (\ref{stationary_continuity}) becomes $\nabla(\nabla\theta-e\mathbf{A}/\hbar c)=0$.  It further simplifies in the Coulomb gauge $\nabla\cdot\mathbf{A}=0$, becoming the Laplace equation $\nabla^{2}\theta=0$.  The latter also follows from imposing the Coulomb gauge $\nabla\cdot\mathbf{A}=0$ in Eq.(\ref{A_large_rho}).  The relevant solution to the Laplace equation $\nabla^{2}\theta=0$ that is consistent with the condition of single-valuedness (\ref{single-valuedness}) is $\theta=l\varphi$.

These considerations motivate the following ansatz for the solution to the GL equations (\ref{GL_equation_Madelung})-(\ref{Ampere's_law_Madelung}) for all $\rho$:
\begin{equation}
\label{conjecture}
\Psi=\sqrt{n_{0}}R(\rho)e^{il\varphi},~~~~~\mathbf{A}=\frac{\hbar cl}{e}\frac{1-F(\rho)}{\rho}\mathbf{e}_{\varphi}.
\end{equation}
Here $R^{2}(\rho)$ is the dimensionless density satisfying the boundary conditions $R(\infty)=1$ and $R(0)=0$.  The latter is a consequence of the single-valuedness of the wave function: as the $z$-axis $\rho=0$ is approached along a fixed $\varphi$ direction, the wave function in (\ref{conjecture}) must approach a $\varphi$-independent limit that can only be zero.  

The dimensionless function $F(\rho)$ entering the expression for the vector potential in Eq.(\ref{conjecture}) satisfies the boundary conditions $F(\rho\rightarrow\infty)\rightarrow 0$ and $F(\rho\rightarrow 0)\rightarrow 1$ (to prevent the $\rho=0$ singularity of the vector potential).  It also determines the behavior of the flow velocity (\ref{flow_velocity}):  
\begin{equation}
\label{vortex_velocity}
\mathbf{v}=\frac{\hbar l}{m}\frac{F(\rho)}{\rho}\mathbf{e}_{\varphi}.
\end{equation} 
The sign of the integer $l$, the topological quantum number, distinguishes clockwise vs counterclockwise circulation of supercurrents around the flux line.
 
Since both the density and flow velocity depend only on $\rho$ while the flow velocity is tangential, $\mathbf{v}\propto\mathbf{e}_{\varphi}$, the continuity equation (\ref{stationary_continuity}) is automatically satisfied.

The function $F(\rho)$ additionally determines the behavior of the magnetic field
\begin{equation}
\label{magnetic_field}
\mathbf{H}=\nabla\times\mathbf{A}=H(\rho)\mathbf{e}_{z}, ~~H(\rho)=-\frac{\hbar cl}{e}\frac{1}{\rho}\frac{dF}{d\rho}
\end{equation}
directed along the $z$-axis.  The expression for the field and the boundary conditions $F(0)=1$ and $F(\infty)=0$ are consistent with the flux quantization (\ref{flux_quantization}) that can now be established bypassing Stokes's theorem:
\begin{equation}
\label{flux_quantization_again}
\int Hdxdy=-\frac{2\pi\hbar c}{e}l\int_{0}^{\infty}\rho d\rho\frac{1}{\rho}\frac{dF}{d\rho}=\frac{2\pi\hbar c}{e}l.
\end{equation}  
Substituting Eqs.(\ref{conjecture}), (\ref{vortex_velocity}) and (\ref{magnetic_field}) into the GL equations (\ref{GL_equation_Madelung}) and (\ref{Ampere's_law_Madelung}), measuring length in units of the coherence length $\xi$, and employing the definitions (\ref{parameters}), we arrive at a system of nonlinear second-order differential equations due to Abrikosov \cite{Abrikosov} for the two unknown functions $R(\rho)$, and $F(\rho)$:
\begin{eqnarray}
\label{A1}
R'' + \frac{1}{\rho}R' - l^2\left(\frac{F}{\rho}\right)^2R-R(R^2-1) &=& 0,\nonumber\\
R(0) = 0,~~~R(\infty) &=& 1,
\end{eqnarray}
\begin{equation}
\label{A2}
F'' - \frac{1}{\rho}F' - \frac{1}{\kappa^2}R^2F = 0,~~ F(0)=1,~~~F(\infty) = 0
\end{equation} 
where the prime is shorthand for the derivative with respect to $\rho$.

\section{Bogomol'nyi method:  completing the square}

The simplest way to arrive at the SB equations is to rewrite the expression for the free energy (\ref{GL_Madelung}) per unit length in terms of the functions $R(\rho)$ and $F(\rho)$ by using Eqs.(\ref{conjecture}), (\ref{vortex_velocity}) and (\ref{magnetic_field}).  Continuing measuring the length in units of the coherence length $\xi$ (\ref{parameters}), the dimensionless free energy per unit length acquires the form
\begin{eqnarray}
\label{dimensionless_fenergy_per_unit_length}
f = \frac{m}{\pi \hbar^2n_0}\frac{\mathcal{F}}{L_z} &=&  \int_{0}^{\infty} \rho d\rho \Bigg\{(R')^2 + l^2\left(\frac{RF}{\rho}\right)^2\nonumber\\
 &+& \frac{1}{2}(R^2-1)^2 + \kappa^2l^2\left(\frac{F'}{\rho}\right)^2\Bigg\}
\end{eqnarray}
where $L_{z}$ is the system size in the $z$-direction.  It is straightforward to verify that variational minimization of this expression with respect to $R$ and $F$ recovers the Abrikosov equations (\ref{A1}) and (\ref{A2}).  

In the spirit of the Bogomol'nyi treatment \cite{Bogo}, the minimization can be carried out differently by grouping the first two integrands, the last two integrands in (\ref{dimensionless_fenergy_per_unit_length}), and completing the square within each group:
\begin{eqnarray}
\label{dimensionless_fenergy_per_unit_length_Bogo1}
f& =&  \int_{0}^{\infty} \rho d\rho \Bigg\{ \left(R' \pm \frac{|l|FR}{\rho}\right)^{2} \mp 2|l| \frac{R'RF}{\rho}\nonumber\\
 &+& \left[\frac{\kappa |l| F'}{\rho} \pm \frac{1}{\sqrt{2}}(R^2-1)\right]^2 \mp \sqrt{2}\kappa |l|\frac{F'(R^2-1)}{\rho}\Bigg\}.\nonumber\\
\end{eqnarray}
Employing the boundary conditions (see Eqs.(\ref{A1}) and (\ref{A2})), the last term in Eq.(\ref{dimensionless_fenergy_per_unit_length_Bogo1}) can be modified via integration by parts into
\begin{equation}
\label{by_parts}
\int_{0}^{\infty} \rho d\rho  \left[\frac{F'(R^2-1)}{\rho}\right]= 1 - 2\int_{0}^{\infty}\rho d\rho \left(\frac{R'RF}{\rho}\right)
\end{equation}
which is then substituted into Eq.(\ref{dimensionless_fenergy_per_unit_length_Bogo1}):
\begin{eqnarray}
\label{dimensionless_fenergy_per_unit_length_Bogo2} 
f &=& \mp \sqrt{2}\kappa|l| +  \int_{0}^{\infty} \rho d\rho \Bigg\{ \left(R' \pm \frac{|l|FR}{\rho}\right)^{2}\nonumber\\
 &+& \left[\frac{\kappa |l| F'}{\rho} \pm \frac{1}{\sqrt{2}}(R^{2}-1)\right]^{2}\Bigg\}\nonumber\\
 &\mp& 2|l|(1-\sqrt{2}\kappa)\int_{0}^{\infty}\rho d\rho \left(\frac{R'RF}{\rho}\right). 
\end{eqnarray}
While the first two integrands are positively defined, the last one is of indefinite sign with the exception of the special case $\kappa=1/\sqrt{2}$, called the Bogomol'nyi point, where it is zero.  In this case, employing the argument that the presence of a defect raises the energy singles out the lower signs in Eq.(\ref{dimensionless_fenergy_per_unit_length_Bogo2}), implying that the dimensionless free energy per unit length has the form  
\begin{eqnarray}
\label{dimensionless_fenergy_per_unit_length_Bogo3}
f = |l| &+&  \int_{0}^{\infty} \rho d\rho \Bigg\{ \left(R' - \frac{|l|FR}{\rho}\right)^{2}\nonumber\\
 &+& \frac{1}{2}\left[\frac{|l|F'}{\rho} - (R^{2}-1)\right]^{2}\Bigg\}.
\end{eqnarray}
Then the free energy per unit length is minimized at $f=|l|$ for the terms in the integrand vanishing, giving the SB equations:
\begin{equation}
\label{B1}
R' = \frac{|l|FR}{\rho},
\end{equation}
\begin{equation}
\label{B2}
\frac{|l|F'}{\rho} = R^{2}-1.
\end{equation}
It is straightforward to verify that differentiating both sides of the first SB equation (\ref{B1}) and employing Eqs.(\ref{B1}) and (\ref{B2}) one obtains Abrikosov's first equation (\ref{A1}).  Likewise, multiplying both sides of the second SB equation (\ref{B2}) by $\rho$ and then differentiating, recovers Abrikosov's second equation (\ref{A2}) at $\kappa=1/\sqrt{2}$.   

A simple consequence of the first-order SB equations (\ref{B1}) and (\ref{B2}) is that $R(\rho)$ is a monotonically increasing function while $F(\rho)$ is a monotonically decreasing function, something that is not immediately obvious from the second-order Abrikosov equations (\ref{A1}) and (\ref{A2}). 

Additionally, the lower signs version of the expression for the free energy per unit length (\ref{dimensionless_fenergy_per_unit_length_Bogo2}) can be combined with the SB equations (\ref{B1}) and (\ref{B2}) to obtain an expression for $f$ in the vicinity of $\kappa=1/\sqrt{2}$:
\begin{equation}
\label{vicinity}
f=|l|\left [1+(\sqrt{2}\kappa-1)\left (1-2\int_{0}^{\infty}R'RFd\rho\right )\right ].
\end{equation}
Since $R(\rho)$ is a monotonically increasing function, the integral in Eq.(\ref{vicinity}) is a positive $|l|$-dependent number.  To determine whether for $\kappa$ close to $1/\sqrt{2}$ the free energy per unit length (\ref{vicinity}) is an increasing or decreasing function of $\kappa$, numerical evaluation is necessary.  Such an evaluation has been carried out for $|l|=1$ \cite{numerical} and  concluded that for $\kappa$ close to $1/\sqrt{2}$ the free energy per unit length (\ref{vicinity}) is an increasing function of $\kappa$.  There are no physical reasons to believe that this conclusion would change for general $l$.

\section{Sarma's operator method}

Following Ref.\cite{de Gennes}, let us introduce the operator of kinetic momentum 
\begin{equation}
\label{kinetic_momentum}
\mathbf{\hat\Pi} = -i\hbar\nabla - \frac {e}{c}\mathbf{A}
\end{equation}
and assume that the magnetic field is pointing in the $z$ direction, $\mathbf{H}=H\mathbf{e}_{z}$.  Then the vector potential only has $x$ and $y$ components and both the vector potential $\mathbf{A}$ and the wave function $\Psi$ only depend on $x$ and $y$.  In this case the GL equations (\ref{GL_equation}) and (\ref{Ampere's_law}) can be rewritten as
\begin{equation}
\label{GL_equation_Sarma}
\left[ \frac{1}{2m} \left ( \hat\Pi_x^2 + \hat\Pi_y^2\right) + \frac{ms^2}{n_0}(|\Psi|^2-n_0)\right]\Psi = 0,
\end{equation}
\begin{equation}
\label{Ampere's_law_Sarma}
\nabla \times \mathbf{H} = \frac{2\pi e}{mc}\left[\Psi^{*} \mathbf{\hat\Pi}\Psi + \Psi \left(\mathbf{\hat\Pi}\Psi\right)^* \right].
\end{equation}
Introducing the operators
\begin{equation}
\label{raising_lowering}
\hat\Pi^{\pm} = \hat\Pi_{x} \pm i\hat\Pi_{y},
\end{equation}
the square of the kinetic momentum operator (\ref{kinetic_momentum}) entering the first GL equation (\ref{GL_equation_Sarma}) can be written as
\begin{equation}
\label{kinetic_momentum_square}
\hat\Pi_{x}^{2} + \hat\Pi_{y}^{2}=\hat{\Pi}^{\mp}\hat{\Pi}^{\pm}\pm\frac{e\hbar}{c}H.
\end{equation}
Limiting ourselves to particular (Sarma) solutions of the form
\begin{equation}
\label{S1}
\hat{\Pi}^{\pm}\Psi=0
\end{equation}
simplifies the first GL equation (\ref{GL_equation_Sarma}) to
\begin{equation}
\label{S2}
\pm\frac{e\hbar}{2mc}H+\frac{ms^{2}}{n_{0}}(|\Psi|^{2}-n_{0})=0.
\end{equation} 
These are the SB equations in their most general and complete form: Eqs.(\ref{S1}) are first-order differential equations in the wave function while Eqs.(\ref{S2}) are first-order differential equations in the vector potential.  At this point we observe that de Gennes \cite{de Gennes} only worked with upper sign solutions to  Eqs.(\ref{S1}) and (\ref{S2}).  On the other hand, the authors of Refs. \cite{recent1,recent3} focused on the lower sign solutions to Eqs.(\ref{S1}) and (\ref{S2}).  We stress that both sign solutions are legitimate and must be kept.  

In order to establish the condition of consistency of Eqs.(\ref{S1}) and (\ref{S2}) with Amp\`ere's law (\ref{Ampere's_law_Sarma}) the latter can be rewritten as 
\begin{equation}
\label{complex_Ampere}
\frac{\partial H}{\partial y}\mp i\frac{\partial H}{\partial x}=\frac{2\pi e}{mc}\left [\Psi^{*}\hat{\Pi}^{\pm}\Psi+\Psi(\hat{\Pi}^{\mp}\Psi)^{*}\right ]
\end{equation}    
With the upper sign in Eqs.(\ref{S1}) and (\ref{S2}) chosen, Amp\`ere's law (\ref{complex_Ampere}) becomes
\begin{equation}
\label{upper_Ampere}
\frac{\partial H}{\partial y}- i\frac{\partial H}{\partial x}=\frac{2\pi e}{mc}\Psi\left (\hat{\Pi}^{-}\Psi\right )^{*}.
\end{equation}
Its left-hand side can be independently computed from Eq.(\ref{S2}):
\begin{eqnarray}
\label{independent1}
\frac{\partial H}{\partial y} - i\frac{\partial H}{\partial x}& =& - \frac{2m^{2}s^{2}c}{n_0e\hbar}\Bigg\{\Psi^{*}\left(\frac{\partial \Psi}{\partial y} - i\frac{\partial \Psi}{\partial x}\right)\nonumber\\
 &+& \Psi\left(\frac{\partial \Psi^{*}}{\partial y} - i\frac{\partial \Psi^{*}}{\partial x}\right)\Bigg\}.
\end{eqnarray}
The upper sign SB equation (\ref{S1}) can be explicitly written as
\begin{equation}
\label{explicitS1}
\frac{\partial \Psi}{\partial y} - i\frac{\partial \Psi}{\partial x} = \frac{e}{\hbar c}\left(A_x+iA_y\right)\Psi
\end{equation}  
and then substituted into Eq.(\ref{independent1}) with the result
\begin{equation}
\label{independent2}
\frac{\partial H}{\partial y} - i\frac{\partial H}{\partial x} =  \frac{2m^{2}s^{2}c}{n_{0}e\hbar^{2}}\Psi\left(\hat\Pi^{-}\Psi\right)^{*}
\end{equation}
Consulting with the definitions (\ref{parameters}) we see that Eqs.(\ref{upper_Ampere}) and (\ref{independent2}) are consistent with each other only at the Bogomol'nyi point $\kappa=1/\sqrt{2}$.

A very similar analysis can be carried out with the lower sign SB equations (\ref{S1}) and (\ref{S2}) to show again that at the Bogomol'nyi point $\kappa=1/\sqrt{2}$ they reduce to the GL equations (\ref{GL_equation_Sarma}) and (\ref{Ampere's_law_Sarma}).

In order to clarify the physical difference between the upper and lower sign solutions of the SB equations (\ref{S1}) and (\ref{S2}),  it is instructive to see how they reduce to Eqs.(\ref{B1}) and (\ref{B2}) that rely on the presence of a topological defect of quantized magnetic flux (\ref{flux_quantization}) and cylindrical symmetry of the magnetic field and density distributions.  

Once again, let us start with the upper sign in Eqs.(\ref{S1}) and (\ref{S2}).  Substituting $|\Psi|^{2}=n_{0}R^{2}(\rho)$, Eq.(\ref{conjecture}), the expression for the magnetic field (\ref{magnetic_field}) into Eq.(\ref{S2}) and measuring the length in units of the coherence length $\xi$ (\ref{parameters}) we find $lF'/\rho=R^{2}-1$ which agrees with Eq.(\ref{B2}) for $l>0$.  Likewise, substituting the expressions for the wave function and the vector potential (\ref{conjecture}) into Eq.(\ref{explicitS1}), and performing differentiations in the left-hand side in Eq.(\ref{explicitS1}), we find $R'=lFR/\rho$ which agrees with Eq.(\ref{B1}) for $l>0$.  It is straightforward to verify that a very similar calculation involving lower sign versions of Eqs.(\ref{S1}) and (\ref{S2}) reproduces Eqs.(\ref{B1}) and (\ref{B2}) for $l$ negative.  So in the particular case of the Abrikosov flux line the two sign solutions to the SB equations (\ref{S1}) and (\ref{S2}) correspond to clockwise or counterclockwise directions of supercurrents circulating around the flux line.   

\section{Combined procedure:  Bogomol'nyi's method employing Sarma's operators}

The derivation of the SB equations by the Bogomol'nyi method (Section IV) relied on the existence of a topological defect carrying quantized magnetic flux (\ref{flux_quantization}) with cylindrically symmetric distributions of the density and magnetic field.  We now present a more general derivation that does not employ cylindrical symmetry.  The procedure parallels Bogomol'nyi's original derivation \cite{Bogo} combining completing the square within the integrand of the free-energy functional with Sarma's operator method.  We emphasize that flux quantization will still be needed.

Assuming that the magnetic field is pointing in the $z$ direction, and employing the fact that the operators $\hat{\Pi}_{x,y}$ (\ref{kinetic_momentum}) are Hermitian, the GL free energy per unit length can be written as:
\begin{eqnarray}
\label{BS1}
\frac{\mathcal{F}}{L_{z}}=\int \Bigg\{ \frac{1}{2m}\Psi^{*}\left (\hat{\Pi}_{x}^{2}+\hat{\Pi}_{y}^{2}\right )\Psi
&+&\frac{ms^{2}}{2n_{0}}(|\Psi|^{2}-n_{0})^{2}\nonumber\\
&+&\frac{H^{2}}{8\pi} \Bigg\} dxdy. 
\end{eqnarray}     
Utilizing the identity (\ref{kinetic_momentum_square}) and the property that the operators $\hat{\Pi}^{\pm}$ (\ref{raising_lowering}) are Hermitian conjugates of each other, Eq.(\ref{BS1}) can be further rewritten as
\begin{eqnarray}
\label{BS2}
\frac{\mathcal{F}}{L_{z}}&=&\int \Bigg\{ \frac{1}{2m}|\hat{\Pi}^{\pm}\Psi|^{2}\pm\frac{e\hbar}{2mc}H|\Psi|^{2}\nonumber\\
&+&\frac{ms^{2}}{2n_{0}}(|\Psi|^{2}-n_{0})^{2}+\frac{H^{2}}{8\pi} \Bigg\} dxdy. 
\end{eqnarray} 
Combing the last two integrands to complete the square and rearranging we find
\begin{eqnarray}
\label{BS3}
\frac{\mathcal{F}}{L_{z}}&=&\int \Bigg\{ \frac{1}{2m}|\hat{\Pi}^{\pm}\Psi|^{2}\pm\left (\frac{e\hbar}{2mc}-\sqrt{\frac{ms^{2}}{4\pi n_{0}}}\right )H|\Psi|^{2}\nonumber\\
&+&\frac{n_{0}}{2ms^{2}}\left [\frac{ms^{2}}{n_{0}}\left (|\Psi|^{2}-n_{0}\right )\pm\sqrt{\frac{ms^{2}}{4\pi n_{0}}}H\right]^{2}\nonumber\\
&\pm& n_{0}\sqrt{\frac{ms^{2}}{4\pi n_{0}}}H \Bigg\}dxdy.
\end{eqnarray} 
While the first and third integrands are positively defined, the second one is of indefinite sign with the exception of the special case when the coefficient in front of $H|\Psi|^{2}$ vanishes.  Employing the definitions (\ref{parameters}) it is then straightforward to verify that this corresponds to $\kappa=1/\sqrt{2}$, the Bogomol'nyi point.  

The last term in Eq.(\ref{BS3}) proportional to the magnetic flux over the $xy$ plane can be dealt with conclusively only if it is a conserved quantity, in other words, the magnetic flux is quantized according to Eq.(\ref{flux_quantization}).  

As a result, at the Bogomol'nyi point $\kappa=1/\sqrt{2}$ the free energy per unit length is minimized at $\mathcal{F}/L_{z}=(\pi\hbar^{2}n_{0}/m)|l|$ which is in agreement with Eq.(\ref{dimensionless_fenergy_per_unit_length_Bogo3}).  Additionally, the magnetic field configuration and density profile is described by the SB equations (\ref{S1}) and (\ref{S2}). 

\section{Conclusion}

To summarize, at the Ginzburg-Landau parameter $\kappa=1/\sqrt{2}$ the second-order nonlinear Ginzburg-Landau equations can be reduced to first-order nonlinear Sarma-Bogomol'nyi equations (\ref{S1}) and (\ref{S2}).  This can be accomplished either by Sarma's method \cite{de Gennes} or by Bogomol'nyi's method \cite{Bogo}.  While at the heart of both methods lies the same operator identity (\ref{kinetic_momentum_square}), relevant details of the procedures are different.  

The Bogomol'nyi approach consists in direct and unorthodox minimization of the Ginzburg-Landau free energy functional that can be completed only if the magnetic flux is quantized.  

On the other hand, in Sarma's method one is dealing with the Ginzburg-Landau equations directly and the only assumption made is that the magnetic field is pointing in the $z$ direction.  Therefore, Sarma's method is more general.  The implication is that the Sarma-Bogomol'nyi equations also apply to situations when the magnetic flux is not quantized.  In fact, these equations were first employed in exactly such a situation \cite{de Gennes} to show that at the value of the Ginzburg-Landau parameter $\kappa=1/\sqrt{2}$ the surface energy of the domain wall separating normal and superconductive phases has zero energy.  Such a domain wall is not characterized by quantized magnetic flux.    

Our conclusions may also be relevant in other field theories where Bogomol'nyi equations are encountered \cite{Manton,Weinberg}.  Indeed, if such equations can be derived at particular combinations of coupling constants via the ingenious energy minimization procedure, it seems plausible they can also be deduced from second-order Euler-Lagrange field equations.  If this is the case, the equations would apply more generally and may be a source of a variety of new solutions beyond topological defects.  We are planning to pursue this line of inquiry in the future. 

\section{Acknowledgements}

We are grateful to E. Babaev, J.-F. Joanny and E. Raphael for informative correspondence and to J. P. Straley for critical comments.

\end{document}